\documentclass[showpacs,amsmath,
twocolumn, 
aps,prl,superscriptaddress]{revtex4}
\usepackage{graphicx}
\usepackage{dcolumn}

\begin{document}

\title{Current-Induced Polarization and the Spin Hall Effect at Room Temperature}

\author{N. P. Stern}
\author{S. Ghosh}
\affiliation{Center for Spintronics and Quantum Computation,
University of California, Santa Barbara, CA 93106 USA}
\author{G. Xiang}
\author{M. Zhu}
\author{N. Samarth}
\affiliation{Department of Physics, Pennsylvania State University,
University Park, Pennsylvania 16802 USA}
\author{D. D. Awschalom}
\email[Electronic address: ]{awsch@physics.ucsb.edu}
\affiliation{Center for Spintronics and Quantum Computation,
University of California, Santa Barbara, CA 93106 USA}

\date{\today}

\begin{abstract}
Electrically-induced electron spin polarization is imaged in
$n$-type ZnSe epilayers using Kerr rotation spectroscopy.  Despite
no evidence for an electrically-induced internal magnetic field,
current-induced in-plane spin polarization is observed with
characteristic spin lifetimes that decrease with doping density. The
spin Hall effect is also observed, indicated by an
electrically-induced out-of-plane spin polarization with opposite
sign for spins accumulating on opposite edges of the sample.  The
spin Hall conductivity is estimated as $3 \pm 1.5$
$\Omega^{-1}m^{-1}/|e|$ at 20 K, which is consistent with the
extrinsic mechanism.  Both the current-induced spin polarization and
the spin Hall effect are observed at temperatures from 10 K to 295
K.

\end{abstract}


\maketitle

The ability to manipulate carrier spins in semiconductors through
the spin-orbit (SO) interaction is one of the primary motivations
behind the field of spintronics.  SO coupling provides a mechanism
for the generation and manipulation of spins solely through electric
fields \cite{Edelstein:1990, Kato:2004a, Kato:2004b}, obviating the
need for applied magnetic fields.  Much of the recent interest in
the consequences of SO coupling in semiconductors surrounds the
production of a transverse spin current from an electric current,
known as the spin Hall effect.  Though predicted three decades ago
\cite{Dyakonov:1971}, the first experimental observations of the
spin Hall effect have appeared only recently \cite{Kato:2004c,
Wunderlich:2005, Sih:2005}.  Subsequent work into the spin Hall
effect has addressed the importance of extrinsic or intrinsic
mechanisms of the spin Hall conductivity \cite{Murakami:2003,
Sinova:2004, Engel:2005, Sih:2005}, the nature of spin currents
\cite{Shi:2006, Sih:2006a}, and the potential ability both to
produce and to detect spin Hall currents using only electric fields
\cite{Erlingsson:2005, Valenzuela:2006}.

Previous experiments showing electrical generation of spin
polarization in semiconductors through SO coupling have been
performed at cryogenic temperatures in GaAs, the archetypical III-V
zincblende semiconductor.  In contrast, the wide band gap and long
spin coherence times of II-VI semiconductors allow many spin-related
effects to persist to higher temperatures than typically observed in
the GaAs system \cite{Malajovich:1999}.  Many of the effects of SO
coupling on the electrical manipulation of spin polarization have
not been studied in detail in these compounds.  In ZnSe, the
extrinsic SO parameter $\lambda_{ZnSe} = 1.06$ $e${\AA}$^2$, as
calculated from an extended Kane model, is five times less than that
in GaAs, with $\lambda_{GaAs} = 5.21$ $e${\AA}$^2$ \cite{Engel:2005,
Winkler:2003}.  Despite weaker SO coupling, large extrinsic SO
skew-scattering has been observed in the anomalous Hall effect in
magnetically doped ZnSe \cite{Cumings:2006}.  In this Letter we
optically measure electrically-induced spin polarization in ZnSe
epilayers that persists to room temperature.  We observe in-plane
current-induced spin polarization (CISP) in ZnSe with n-doping
ranging over two orders of magnitude and out-of-plane
electrically-induced spin accumulation at the edges of an etched
channel, providing evidence for the extrinsic spin Hall effect.
Unlike in previous studies of CISP and the spin Hall effect, both
phenomena are measured at 300 K, demonstrating the electrical
generation and routing of spins in semiconductors at room
temperature.

A series of 1.5 $\mu$m thick $n$-type Cl-doped ZnSe epilayer samples
with room temperature carrier concentrations $n = 5 \times 10^{16}$
cm$^{-3}$, $9 \times 10^{17} $ cm$^{-3}$, and $9 \times 10^{18}$
cm$^{-3}$ are grown by molecular beam epitaxy on semi-insulating
(001) GaAs substrates.  Perpendicular channels of width $ w = 100$
$\mu$m and length $l = 235$ $\mu$m are patterned along [110] and
[1$\overline{1}$0] directions of the ZnSe epilayers, allowing an
electric field $E$ to be applied along both the crystal axes.  A
voltage is applied across the device, with the effective $E$
calculated from the measured temperature-dependent resistivity and
current to eliminate the effect of contact resistance.

The samples are mounted in the variable temperature insert of a
magneto-optical cryostat.  Kerr rotation (KR) is measured in the
Voigt geometry, with an in-plane applied magnetic field $B$
perpendicular to the laser propagation direction (Fig. 1a).  150-fs
pulses from a 76-MHz mode-locked Ti:sapphire laser are
frequency-doubled and split into a circularly polarized pump and a
linearly polarized probe beam with powers of 1.2 mW and 400 $\mu$W,
respectively.  The Kerr rotation angle $\theta_K$ of the
polarization axis of the reflected probe beam measures the
projection of electron spin polarization along the propagation
direction \cite{Baumberg:1994}. Time-resolved KR measurements have
found the electron g-factor to be $g = 1.1$ and the spin coherence
time to decrease with increasing $n$-doping, with spin coherence
times of 50 ns, 20 ns, and 0.5 ns for the $n = 5 \times 10^{16}$
cm$^{-3}$, $9 \times 10^{17} $ cm$^{-3}$, and $9 \times 10^{18}$
cm$^{-3}$ samples, respectively, at $T = 5$ K and $B = 0$ T
\cite{Malajovich:2000}.

\begin{figure}[b]\includegraphics{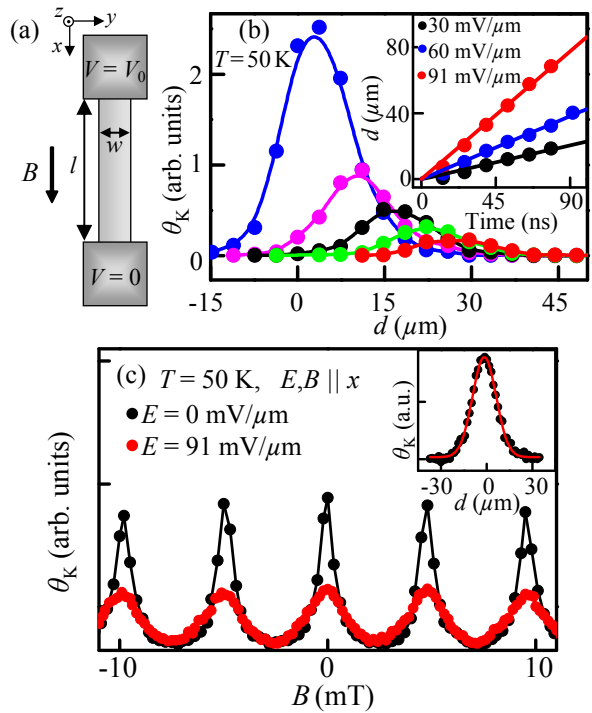}\caption{\label{fig1}
(a) Schematic of the device geometry for spin drag and CISP
measurement.  $B \parallel E$ and the probe beam is incident along
$-z$.  (b)  Spatial profiles of the optically injected spin packet
extracted from Fourier transforms of $\theta_K (B)$ at $E$ = 60
mV/$\mu$m\cite{Kikkawa:1999} for $\Delta t$ = 13.1 ns (blue),
$\Delta t$ = 26.2 ns (magenta), $\Delta t$ = 39.3 ns (black),
$\Delta t$ = 52.4 ns (green), and $\Delta t$ = 65.5 ns (red).
Gaussian fits at each laser repetition ($\Delta t$ = 13.1 ns) give
the center position of the packet as a function of time (inset).
(c) KR from the $n = 5 \times 10^{16}$ cm$^{-3}$ sample at $E$ = 0
mV/$\mu$m (black) and $E$ = 91 mV/$\mu$m (red).  The KR peaks remain
centered at $B=0$, showing no evidence of $B_{int}$.  The inset
shows the Gaussian spatial profile of the optically injected spin
packet, with a width of 15.5 $\mu$m. }\end{figure}

In order to characterize the response of electron spins in ZnSe to
applied electric fields, we perform spatially resolved KR
measurements. In this pump-probe technique, the beams are normally
incident on the sample and focused to a 15 $\mu$m spot (Fig. 1c
inset).  The relative separation ($d$) of the pump and probe is
varied in the direction of the electric field, and the KR of the
probe measures the electron spin polarization injected by the pump
along the $z$-axis. Figure 1b follows the optically injected spin
packet as it is dragged along the channel by a DC electric field of
60 mV/$\mu$m in the $n = 5 \times 10^{16}$ cm$^{-3}$ sample.
Extracting the drift velocity from the center of the Gaussian spin
packets allows an estimate of the spin mobility of $\mu_s$ $= 89 \pm
14$ cm$^2 / $Vs.  This is 20 times less than that measured in GaAs
\cite{Kikkawa:1999} and over an order of magnitude smaller than the
ZnSe electron mobility $\mu_e$ = 1440 cm$^2 / $Vs at $T = 50$ K for
this sample.

    Experiments in GaAs have shown that an internal magnetic field $B_{int}$ acts on electrons accelerated by an electric field, which has been attributed to inversion asymmetry \cite{Kato:2004a, Kikkawa:1999, Crooker:2005,  Sih:2006b}.  KR as a function of $B$ with fixed spatial ($d = 0$) and temporal (13.1 ns) pump-probe separation is shown in Fig. 1c. This signal is periodic in $B$ and symmetric about $B = 0$, making it a very sensitive probe for detecting $B_{int}$ \cite{Kikkawa:1999, Kato:2004a}. The KR signal remains centered at $B = 0$ as we increase $E$, showing no evidence of a $B_{int}$ in the $n = 5 \times 10^{16}$ cm$^{-3}$ and $n = 9 \times 10^{17}$ cm$^{-3}$ samples along either the [110] or [1$\overline{1}$0] channel.  The spin coherence time of the $n = 9 \times 10^{18}$ cm$^{-3}$ sample is too short to observe KR at 13.1 ns temporal separation, but no evidence of electrically-induced spin precession from a $B_{int}$ is observed using time-resolved KR with $B = 0$ \cite{Kato:2004a}.  These measurements provide an upper bound for the internal magnetic field of $0.1$ mT with an $E$ = 91 mV/$\mu$m.  The lack of any observable $B_{int}$ in ZnSe can be attributed to the weaker spin-orbit coupling in ZnSe and the minimal in the epilayers.

For optical detection of CISP, we block the pump and measure static
KR with probe energy tuned near the maximum of the KR signal,
typically around 2.8 eV at 50 K.  The KR is detected with a lock-in
synched to a 2-kHz applied square wave electric field $E$.  Typical
magnetic field sweeps of KR at $T = 50$ K are shown for each sample
in Fig. 2 with $B \parallel E$.  The characteristic odd-Lorentzian
shape is indicative of spins generated in-plane and perpendicular to
$E$ \cite{Kato:2004b}. The data are modeled as spins generated along
the $y$ direction, with a background subtracted, and are fit to
$\theta_{el}  \omega_L \tau / [(\omega_L \tau)^2 + 1]$, where
$\theta_{el}$ is the KR amplitude and $\omega_L = g \mu_B B / \hbar$
is the Larmor precession frequency, $\mu_B$ is the Bohr magneton,
and $\tau$ is the spin coherence time \cite{Kato:2004b}.  We measure
$\theta_{el}$ to be independent of the square wave frequency and
linear with both $E$ and probe power. The trends in $\tau$ between
samples match the trend in spin coherence time
\cite{Malajovich:2000}, but the values are not numerically
identical.   For $n = 5 \times 10^{16}$ cm$^{-3}$ and $n = 9 \times
10^{17}$ cm$^{-3}$, $\tau$ decreases with increasing E, but the $n =
9 \times 10^{18}$ cm$^{-3}$ sample exhibits little change in $\tau$.
CISP has also been observed in other samples of lower doping density
($n \sim 1 \times 10^{16}$cm$^{-3}$), but systematic results are
difficult due to large resistivity.  Further quantitative optical
analysis is performed as in Ref. \onlinecite{Kato:2004b} to estimate
the efficiency of the electrical spin generation  giving
$\theta_{el} \approx 12$ spins $\mu m^{-3}$ at 20 K.  The sign in
the figure corresponds to spins generated along the $+y$ direction
when the electric field is in the $+x$ direction.

\begin{figure}\includegraphics{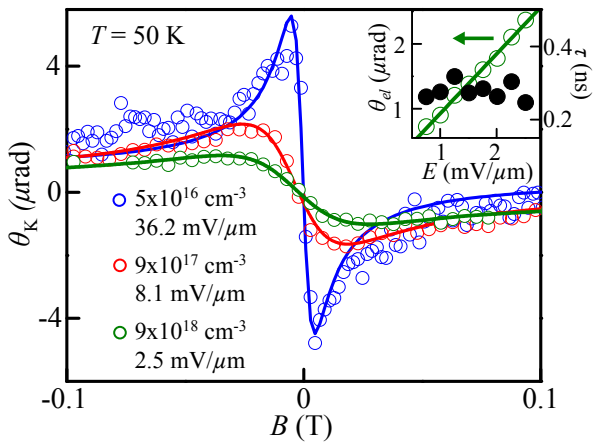}\caption{\label{fig2}
(a) $\theta_K$ as a function of $B$ for $n$-ZnSe at $T$ = 50 K.
Open circles are the data, while the solid lines are fits to the
data as described in the text.  The maximum $E$ that can be applied
to each sample without heating decreases with increasing $n$ due to
lower sample resistances. The inset shows the electric field
dependence of the $\theta_{el}$ and $\tau$ for the $n = 9 \times
10^{18}$ cm$^{-3}$ sample. }\end{figure}

The microscopic origin of CISP is not well-understood
\cite{Kato:2004b, Bernevig:2005}.  In-plane spin generation along
the Rashba spin-orbit field \cite{Edelstein:1990, Culcer:2005} has
been used to explain CISP in two-dimensional electron
\cite{Yang:2005} and hole \cite{Silov:2004} gases.  Following the
same formalism, strain-enhanced inversion asymmetry terms in the
Hamiltonian manifest as $B_{int}$ and could generate the spin
polarization \cite{Kato:2004b, Bernevig:2005}.  In general, the
internal magnetic field strength shows a close correlation to the
amount of strain in GaAs structures \cite{Kato:2004a, Sih:2006b},
but the magnitude of CISP shows little correlation to the strength
of $B_{int}$ \cite{Kato:2004b}. In the current experiment in
$n$-ZnSe, the CISP is comparable in magnitude to that in $n$-GaAs,
even with no observable $B_{int}$.

The spin Hall effect is probed using a low-temperature scanning Kerr
microscope with a spatial resolution of approximately 1 $\mu$m
\cite{Stephens:2003, Kato:2004c, Sih:2005}.  The ZnSe channel is
mounted with $B \perp E$ $(B \parallel y)$ so in-plane CISP does not
precess and is not detected.  No differences in spin accumulation
between the [110] and [1$\overline{1}$0] channel are observed.
Figure 3a shows the geometry for the spin Hall effect measurements,
with the laser propagating along $-z$.  The origin is taken to be
the center of the channel.  Figure 3b shows typical KR data for
scans of $B$ near the edges of the channel at $ y = \pm 48$ $\mu$m
on the $n = 9 \times 10^{18}$ cm$^{-3}$ sample.  The KR curves are
analogous to the Hanle effect, in which an out-of-plane spin
polarization decreases with $B$ due to spin precession
\cite{Kato:2004c}; these data can be fit to a Lorentzian
$\theta_{el}/[(\omega_L \tau)^2 + 1]$.  The opposite sign of the
spin accumulation on each edge of the sample is a signature of the
spin Hall effect.  This phenomenon is also observed in ZnSe with $n
= 8.9 \times 10^{18}$ cm$^{-3}$, but all the results presented here
are from the sample with $n = 9 \times 10^{18}$ cm$^{-3}$ for
brevity.  Observation of the spin Hall effect is highly dependent on
$n$-doping, as no spin Hall signature is measured in samples with
lower $n$.  The growth of higher doped samples is restricted by MBE
conditions.

\begin{figure}\includegraphics{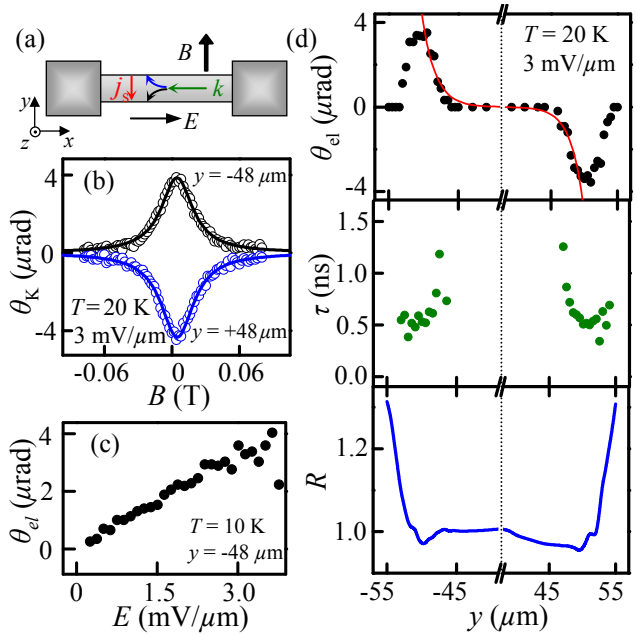}\caption{\label{fig3}
(a)  Schematic showing the measurement geometry for the spin Hall
effect, with $B \parallel y$.  For $E > 0$, $j^{s}_y  < 0$.  (b)
$\theta_K$ (open circles) and fits (lines) at $x = 0$ $\mu$m as a
function of $B$ for $y = - 48$ $\mu$m (black) and $y = +48$ $\mu$m
(blue) at $T$ = 20 K.  (c)  Electric field dependence of the spin
accumulation amplitude $\theta_{el}$.  Above $E$ = 3 mV/$\mu$m the
signal deteriorates due to heating.  (d)  Spatial dependence of the
fit parameters $\theta_{el}$ and $\tau$, as well as the reflectivity
$R$ of the beam (normalized to 1 at $y = 0$), which is used to
monitor the position. }\end{figure}

The amplitude of the spin accumulation $\theta_{el}$ is linear in
$E$ (Fig. 3c), while no appreciable change in $\tau$ is observed
with increasing $E$.  As observed for the spin Hall effect in GaAs,
$\tau$ increases away from the channel edge (Fig. 3d).  The sign and
magnitude of the accumulated spins are found by direct comparison to
CISP in a geometry with $E \parallel B$, which is calibrated by
comparison to time-resolved KR.  At 20 K, the peak spin density near
the edges is aprroximated $n_0 \approx 16$ spins/$\mu$m$^3$, with
spin polarization along $+z$ ($-z$) on the $y = -50 \mu $m ($y = +50
\mu$m) edge for $E >0$ along $x$.  Assuming a simple spin
drift-diffusion model for the accumulation sourced by a spin
current, the profile can be fit by $\theta_{el} = -n_0
\text{sech}(w/2L_s) \text{sinh}(y/L_s)$ \cite{Zhang:2000, Tse:2005,
Kato:2004c}, where $L_s$ is the spin diffusion length (Fig. 3d).
These fits give $L_s = 1.9 \pm 0.2$ $\mu$m at $T = 20$ K.  Ignoring
complications arising from boundary conditions, the spin current
density along $y$ can be written as $|j^{s}_y|  = L_s n_0 / \tau
$\cite{Kato:2004c} and we can calculate the spin Hall conductivity,
$\sigma_{SH} = - j^{s}_y / E_x = 3 \pm 1.5$ $\Omega^{-1}m^{-1}/|e|$
at $T = 20$ K.  Uncertainties in the overall optical calibration
make this only an order-of-magnitude estimate.

The spin Hall conductivity for ZnSe is of comparable magnitude and
of the same sign as that predicted by theory \cite{Engel:2005,
Tse:2006} for GaAs with a dominant extrinsic spin Hall effect.  The
extrinsic spin Hall effect has contributions of differing sign from
both skew scattering and the side jump mechanism.  For the
conditions of Ref. \onlinecite{Kato:2004c}, skew scattering likely
dominates giving $\sigma_{SH} > 0$.  The dominance of skew
scattering should persist in the degenerately $n$-doped ZnSe studied
here since the Fermi energy is well above the conduction band edge
\cite{Tse:2006}.  Intrinsic spin Hall conductivity should have the
opposite sign ($\sigma_{SH} < 0$) \cite{Sinova:2004} and a lower
magnitude \cite{Engel:2005} than measured here; hence, the observed
spin Hall effect in ZnSe is likely extrinsic.

\begin{figure}\includegraphics{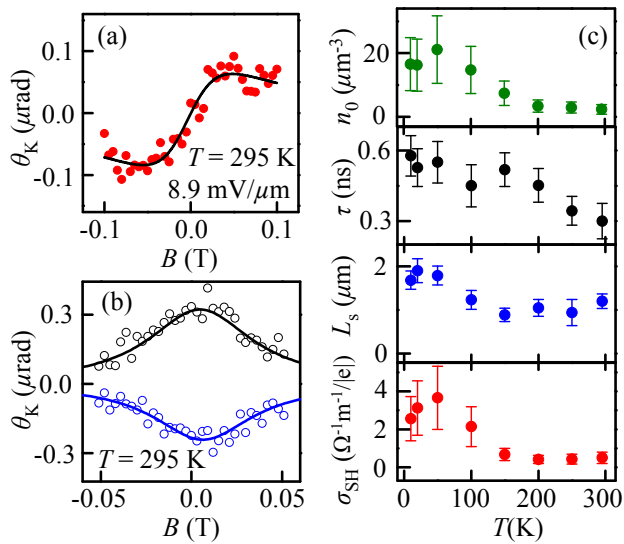}\caption{\label{fig4}
(a)  KR (circles) and fit (line) of CISP at room temperature.
Adjacent-point averaging was done to improve signal-to-noise.  (b)
KR (circles) and fits (lines) of spin Hall polarization at $y = - 48
\mu$m (black) and $y = +48 \mu$m (blue) for $T$ = 295 K.  (c)
Temperature dependence of density $n_0$ (green), coherence time
$\tau$ (black), spin diffusion length $L_s$, and spin Hall
conductivity $\sigma_{SH}$. }\end{figure}

Measurements of both CISP and the spin Hall effect at higher
temperatures show a decrease in the spin coherence time $\tau$ and
the peak spin polarization $n_0$, but both phenomena persist up to
room temperature (Fig. 4 a,b).  Figure 4c shows temperature
dependences of the various parameters discussed above.  The spin
polarization is an order of magnitude weaker at room temperature and
$L_s$ decreases from 1.9 $\mu$m at 20 K to 1.2 $\mu$m at 295 K.  The
estimated spin Hall conductivity decreases to $\sigma_{SH} \approx
0.5$ $\Omega^{-1}m^{-1}/|e|$ at room temperature.

These results demonstrate electrically-induced spin polarization and
the extrinsic spin Hall effect at room temperature in a II-VI
semiconductor.  Despite the absence of a measurable internal field
and the weaker spin-orbit coupling in ZnSe compared to GaAs, these
phenomena remain measurable.  The remarkable ability for
all-electrical spin generation at room temperature suggests that
spin-based logic is technologically feasible in semiconductor
devices.

\begin{acknowledgments}
We thank NSF and ONR for financial support.  N.P.S. acknowledges the
support of the Fannie and John Hertz Foundation.
\end{acknowledgments}

\end{document}